\documentstyle[twoside,12pt]{article}
\normalsize
\def\greaterthansquiggle{\raise.3ex\hbox{$>$\kern-.75em\lower1ex\hbox{$\sim$}}}
\def\lessthansquiggle{\raise.3ex\hbox{$<$\kern-.75em\lower1ex\hbox{$\sim$}}}
\newcommand{\bdi}{\begin{displaymath}}
\newcommand{\edi}{\end{displaymath}}
\newcommand{\bfi}{\begin{figure}}
\newcommand{\efi}{\end{figure}}

\newcommand{\beq}{\begin{equation}}
\newcommand{\eeq}{\end{equation}}

\newcommand{\gaf}{\gamma_{5}}

\newcommand{\beqa}{\begin{eqnarray}}
\newcommand{\eeqa}{\end{eqnarray}}
\newcommand{\no}{\nonumber}

\newcommand{\ra}{\rightarrow}

\newcommand{\ddsla}{\partial\hspace{-4.6pt} /  }

\newcommand{\AAsla}{A\hspace{-5pt}  /  }

\begin{document}
\bibliographystyle{plain}

\begin{titlepage}
\begin{flushleft} 
FSUJ TPI 13/96
\end{flushleft}
\begin{flushright}
September, 1996
\end{flushright}

\vspace{1cm}
\begin{center}
{\Large \bf Charge screening and 
confinement in the massive Schwinger model}\\[1cm]
C. Adam* \\
Friedrich-Schiller-Universit\"at Jena, Theoretisch-Physikalisches Institut \\
Max-Wien Platz 1, D-07743 Jena, Germany
\vfill
{\bf Abstract} \\
\end{center}
Within the framework of Euclidean path integral and mass perturbation theory we
compute the Wilson loop of widely separated external charges for the massive
Schwinger model. From this result we show for arbitrary order mass perturbation
theory that integer external charges are completely screened, whereas for
noninteger charges a constant long-range force remains.

\vfill

$^*)${\footnotesize permanent address: Inst. f. theoret. Physik d. Uni Wien \\
Boltzmanngasse 5, 1090 Wien, Austria \\
email address: adam@pap.univie.ac.at}
\end{titlepage}

\section{Introduction}

The Schwinger model is two-dimensional QED with one fermion flavour. For a
massless fermion the model is exactly soluble and has been studied intensively
in the last decades (\cite{Sc1} -- \cite{ABH}). 
In fact, it is equivalent to the theory of one massive,
free scalar field (Schwinger boson), which may be interpreted as a
fermion-antifermion bound state. When the fermion is massive (massive Schwinger
model), the Schwinger boson turns into an interacting particle that may form
bound states and undergo scattering processes (\cite{Co1} -- \cite{GBOUND}).

Another feature of both massless and massive model is that the perturbative
vacuum is not invariant under large gauge transformations, and, therefore,
instanton-like gauge fields are present and a $\theta$ vacuum has to be
introduced as a new, physical vacuum (\cite{LS1} -- \cite{Adam},
\cite{Co1,MSSM,CJS}).

A further aspect that is present in both models and stimulated some
investigations is confinement (\cite{AAR,CKS,KS1,GKMS}). 
In both models there are no fermions in the
physical spectrum, so "confinement" is realized in a certain sense. When these
confinement properties are further tested by putting widely separated external
probe charges into the vacuum, the two models, however, behave differently. In
the massless model the two charges are completely screened by vacuum
polarization, and the "quark-antiquark" potential approaches a constant for
large distances. This may be interpreted as a dynamical Higgs mechanism, where
the Schwinger boson acts as a massive gauge boson.

In \cite{GKMS} the following behaviour of the massive Schwinger model was shown
to hold in first order mass perturbation theory: As long as the external
charges are integer multiples of the fundamental charge, $g=ne$, $n\in {\rm\bf
N}$, these charges are completely screened as in the massless model. On the
other hand, when $g\ne ne$, the potential between the probe charges rises
linearly for large distances. So "screening" is realized in the massless model,
whereas true confinement takes place in the massive model.

In this article we want to generalize the first order 
result of \cite{GKMS} to arbitrary order mass perturbation theory (mass
perturbation theory is discussed e.g. in \cite{FS1,MSSM,GBOUND}).

\section{String tension from the Wilson loop} 

The Euclidean vacuum functional of the massive model, for general $\theta$, has
the following two equivalent representations
\beqa
Z(m,\theta ) &=&
\sum_{k=-\infty}^{\infty}\int DA_k^\mu D\bar\Psi
D\Psi e^{\int dx[\bar\Psi (i\ddsla -e\AAsla +m)\Psi +\frac{1}{2}F^2 +
\frac{e\theta}{2\pi}F]} \\
&=& \sum_{k=-\infty}^{\infty}\int DA_k^\mu D\bar\Psi
D\Psi e^{\int dx[\bar\Psi (i\ddsla -e\AAsla )\Psi +m\cos\theta\bar \Psi
\Psi +im\sin\theta\bar\Psi \gaf \Psi +\frac{1}{2}F^2 ]}
\eeqa
($F=\frac{1}{2}\epsilon_{\mu\nu}F^{\mu\nu}$, $k\ldots$ instanton number)),
which are related by the chiral anomaly. $Z(m,\theta )$ may be computed within
mass perturbation theory by simply expanding the mass term in (1), see 
\cite{MSSM}. It may be proved that $Z(m,\theta )$ exponentiates (as is, of
course, expected),
\beq
Z(m,\theta )=e^{V\epsilon (m,\theta )}
\eeq
\beq
\epsilon (m,\theta )=m\Sigma\cos\theta +\frac{m^2 \Sigma^2}{4\mu_0^2}(E_+ \cos
2\theta +E_- )+o(m^3)
\eeq
where $V$ is the space-time volume, $\Sigma$ is the fermion condensate of the
massless model, $\mu_0$ is the Schwinger boson mass of the massless model,
$\mu_0^2 =\frac{e^2}{\pi}$, and $E_+ (E_-)$ are some numbers ($E_+ =-8.91,E_-
=9.74$, see \cite{MSSM}); $\epsilon (m,\theta )$ is the vacuum energy density.
A property of $\epsilon (m,\theta )$ that we need in the sequel is the fact
that it is an even function of $\theta$, 
\beq
\epsilon (m,\theta )=\sum_{l=0}^\infty \epsilon_l \cos l\theta ,
\eeq
where the instanton sectors $k=\pm l$ contribute to $\epsilon_l$ (the
$\epsilon_l$, in principle, contain arbitrary orders of $m$).
This property, equ. (5), may be seen most easily from the representation (2) of
the vacuum functional. Indeed, suppose we expand the mass term in (2). There
the perturbation term is $\int dx (m\cos\theta \bar \Psi \Psi + im\sin\theta
\bar \Psi \gaf \Psi)$, and the expansion is about a massless theory with
vanishing vacuum angle, $\theta=0$. As a consequence, parity is conserved in
the theory we are expanding about, and only even powers of $P=\bar\Psi
\gaf \Psi$ may contribute in the perturbation series. Therefore, arbitrary
powers of $\cos\theta$ but only even powers of $\sin\theta$ may occur, which we
wanted to prove.

So let us turn to the determination of the confinement behaviour.
A usual way to investigate confinement is the computation of the string
tension from the Wilson loop. The Wilson loop for a test particle of
arbitrary charge $g=qe$ is defined as (the additional factor $i$ in
Stokes' law is due to our Euclidean conventions, see e.g.
\cite{ABH})
\beq
W_D =\langle e^{ig\int_{\partial D} A_\mu dx^\mu}\rangle = \langle e^{g
\int_D F(x) d^2 x}\rangle =\langle e^{2\pi iq\int_D \nu (x) d^2 x}\rangle
\eeq
where $\nu (x)$ is the Pontryagin index density, $\nu (x)=-\frac{ie}{2\pi} 
F(x)$ and $\nu$ the Pontryagin index. Further $\partial D$ is
the contour of a closed loop and $D$ the enclosed region of space-time.
We are interested in the string tension for very large distances; further
we are able to explicitly separate the area dependence, therefore we set
$D\ra V$ in the sequel.

For the VEV of an exponential the following formula holds,
\beq
\langle e^{2\pi iq\nu }\rangle =\exp 
\Bigl[ \sum_{n=1}^\infty \frac{(2\pi iq)^n}{n!}
\langle \nu^n \rangle_c \Bigr]
\eeq
where $\langle \rangle_c$ denotes the connected part of the $n$-point function.
These VEVs are given by
\beq
\langle \nu^n \rangle_c =V\int dx_2 \ldots dx_n \langle \nu (0) \nu (x_2)
\ldots \nu (x_n)\rangle_c =V(-i)^n \frac{\partial^n}{\partial\theta^n}
\epsilon (m,\theta)
\eeq
as is obvious from the vacuum functional (1). 
Performing the derivatives (8) on the vacuum energy density (5) we have
to separate even ($\sim \cos l\theta$) and odd ($\sim \sin l\theta$)
powers of derivatives. We find for the Wilson loop
\beqa
W&=& \exp \Bigl[V\sum_{n=1}^\infty\frac{(2\pi q)^{2n}}{(2n)!}
\sum_{l=0}^\infty (-1)^n l^{2n}\epsilon_l \cos l\theta 
+V\sum_{n=1}^\infty \frac{(2\pi q)^{2n-1}}{(2n-1)!}
\sum_{l=0}^\infty (-1)^n l^{2n-1} \epsilon_l \sin l\theta \Bigr] \no \\
&=& \exp \Bigl[ V\sum_{l=0}^\infty \epsilon_l \cos l\theta
\sum_{n=1}^\infty (-1)^n \frac{(2\pi ql)^{2n}}{(2n)!} 
+V\sum_{l=l}^\infty \epsilon_l \sin l\theta
\sum_{n=1}^\infty (-1)^n \frac{(2\pi ql)^{2n-1}}{(2n-1)!} \Bigr] \no \\
&=& \exp \Bigl[ V\sum_{l=0}^\infty \epsilon_l \cos l\theta
(\cos 2\pi ql -1) 
 -V\sum_{l=0}^\infty \epsilon_l \sin l\theta \sin 2\pi ql \Bigr] .
\eeqa
The string tension is defined as
\beq
\sigma := -\frac{1}{V} \ln W 
= \sum_{l=0}^\infty \epsilon_l \Bigl( \cos l\theta (1-\cos 2\pi ql) +
\sin l\theta \sin 2\pi ql \Bigr)
\eeq
and may be interpreted as the force between two widely separated probe
charges $g=qe$, where all quantum effects are included.

As indicated in the introduction, we find that, 
whenever the probe charge is an integer multiple of
the fundamental charge, $q \in {\rm\bf N}$, the charges are screened, and
the Wilson loop does not obey an area law. Observe that this result is
{\em exact} !

For noninteger probe charges there is no complete screening and the string
tension may be computed perturbatively,
\bdi
\sigma =m\Sigma \Bigl( \cos\theta (1-\cos 2\pi q)+\sin\theta \sin 2\pi q
\Bigr) + 
\edi
\beq
\frac{m^2 \Sigma^2 E_+}{4\mu_0^2}\Bigl( \cos 2\theta (1-\cos 4\pi q) +\sin
2\theta \sin 4\pi q\Bigr) +o(m^3)
\eeq
showing that in the massive Schwinger model and for noninteger probe charges
there remains a constant force (linearly rising potential) for very
large distances. So in the massive Schwinger model true confinement is
realized instead of charge screening in the general case.

\section{Summary}

As claimed, we succeded in generalizing the first order results of \cite{GKMS}
to arbitrary order mass perturbation theory: integer external probe charges are
completely screened, whereas a linearly rising potential is formed
between widely separated noninteger probe charges. There has been some debate
about this point, therefore we should perhaps add a comment.
From equ. (8) it is obvious that only nontrivial instanton sectors may
contribute to the formation of the string tension. Therefore this string
tension is a strictly nonperturbative phenomenon in the sense of ordinary
(electrical charge $e$) perturbation theory and may not be detected by
conventional perturbative methods. On the other hand, the mass perturbation
theory is an expansion about the true, physical vacuum and takes all the 
nontrivial
structure of the model into account. Therefore, a property that holds for
arbitrary order mass perturbation theory should hold for the model as an exact
property. 

The behaviour of the massless model -- complete charge screening for arbitrary
probe charges -- we find as a trivial byproduct of our result (e.g. equ. (11)).

\section*{Acknowledgement}

The author thanks the members of the Institute of Theoretical Physics of the
Friedrich-Schiller Universit\"at Jena, where this work was done, for their
hospitality.

This work was supported by a research stipendium of the Vienna University.

\end{document}